\documentclass[preprint]{aastex61}
\usepackage{natbib}
\usepackage{amsmath}
\usepackage{amssymb}
\usepackage{graphicx}
\usepackage{epstopdf}
\usepackage{multirow}

\newcommand{\mppc}[1]{\:M_\odot {\rm pc}{}^{-#1}}

\begin{document}
\title{Revisiting the Dark Matter - Comet Shower Connection}
\author{Eric David Kramer, Michael Rowan}
\affil{Department of Physics, Harvard University, Cambridge, MA, 02138}

\begin{abstract}
We revisit the question of whether the observed periodicity of comet impacts on Earth is consistent with Solar oscillation about the Galactic midplane and spiral arm crossings, here in the context of dissipative dark matter models. Consider whether a hypothetical thin dark disk, a signature of these models, is necessary to give the right periodicity, and whether such a dark disk is allowed given kinematic and other observational constraints on the Galaxy's gravitational potential, taking into account recent updates of these limits based on the vertical epicyclic oscillations of the tracer populations. Our analysis contains updated parameters for the Galactic disk, a self-consistent gravitational potential under the Poisson-Jeans equations, and includes prior probabilities from local stellar kinematics and the distribution of Milky Way interstellar gas. Moreover, our analysis also includes radial oscillations and Galactic spiral arm crossings. We find a dark disk explanation for the comet periodicity to be 10 times more likely than a constant average rate model. Moreover, we find that spiral arm crossing is necessary to correctly predict the date of the Chicxulub crater dated to 66 My ago.
\end{abstract}

\section{Introduction}

Many authors have underlined a potential periodicity in the rate of comet impacts on Earth. A possible cause hypothesized to explain this periodicity is the oscillation of the Sun about the galactic midplane. Indeed, this is not a new problem and the possibility has been pursued from a variety of angles \citep{hutalvarez,stothers88,matese,rampino} and the many references therein. 
Accordingly, many authors \citep{stothers98,shaviv,dino} and others suggest the existance of a significant distribution of dark matter in the galactic disk. However, \citet{shaviv16_2} suggests that this disk is too thick to cause comet showers. \citet{matese} also suggests, based on \citet{hf2000}, that the amount of dark matter required to cause periodic comet impacts is ruled out by local stellar kinematics. The goal of this paper is to examine the question of the dark matter-dinosaur connection more carefully, with a global analysis incorporating all of its currently known observable constraints. These include:
\begin{enumerate}
\item A dark disk model motivated by particle physics, and that is also
\item Consistent with the remaining Galactic disk parameters under the Poisson-Jeans equation
\item Recently updated Galactic disk parameters of \citet{paper2} and \citet{mckee}.
\item Inclusion of prior probabilities from kinematic data, taking into account vertical epicyclic oscillations of the star populations from \citet{paper1}
\item Inclusion of prior probabilites from the distribution of interstellar gas from \citet{paper2}.
\item Inclusion of radial oscillations
\item Inclusion of spiral arm crossings
\end{enumerate}
Performing a global analysis including these together will give a more conclusive answer on the question of the dark matter-dinosaur connection. A Poisson-Jeans model was used in \citet{bahcallbahcall}, but this model was not up-to-date. Radial oscillations were included in \citet{shaviv16_2}, but not spiral arm crossings. Lastly, the prior probabilities are important because they allow the evidence of the crater record to be interpreted in the context of the existing literature and constraints. Based on the above analysis, we find that the likelihood ratio between the dark matter cause for the dinosaur extinction and the random cause to be close to 12. Including prior probabilities from stellar kinematics and the distribution of local interstellar gas increases this likelihood ratio to 10. This indicates that this connection should be taken seriously.

\section{Analysis}
\label{sec:analysis}

\subsection{Galactic Mass Model}
We compute the oscillations of the Sun through the galactic plane by integrating the motion of the Sun in the potential computed assuming the mass model described in \citet{paper1,paper2}. As explained in \citet{paper1}, this Bahcall-type model \citet{bahc84b} was obtained by solving the Poisson-Jeans equation for a superposition of isothermal components. The gas parameters used are explained in \citet{paper2}. These were an amalgam of a number of sources. For comparison, the more widely accepted parameters of \citet{mckee} were also used.

\subsection{Crater Record}
We followed the Bayesian analysis of \citet{dino} in comparing the predicted comet rates from oscillations to a constant comet rate. We used the craters dated to the past 250 My from the Earth Impact Database, developed and maintained by the Planetary and Space Science Center of the University of New Brunswick\cite{earthimpact}. We also follow \citet{renne} in correcting the age of the Chicxulub crater from 64.98 $\pm$ 0.05 My to 66.04 $\pm$ 0.05 My\footnote{This may also explain why no $\rm {}^3 He$ flux increase, associated with comet showers, was observed between 63.9 and 65.4 Mya \citep{helium3}.}. We included only craters larger than a certain diameter; we performed the analysis for diameter cutoffs of $D=$ 20 km, 40 km, and 50 km, giving sample sizes of $N_D=$ 26, 13, and 7 craters respectively. Although the 50 km sample is small, according to \citet{shoemaker1988}, it contains craters with a high probability of being caused by comets and not asteroids. The maximum likelihood ratio, however, was found for $D\geq 20$ km.

For an impulse-like disturbance in the tidal force, a comet shower profile was computed by \citet{hutalvarez}. Despite using many updates in the crater record and the Milky Way's mass distribution, we assume the comet shower profile of \citet{hutalvarez} to still be accurate. We therefore convolved this comet shower profile with the predicted tidal force to give the predicted comet rate as a function of age. In the constant rate model, we assumed a constant rate of $N_D=26$, 13, or 7 comets per 250 My.

\subsection{Statistics}
We derived the likelihood for a given model based on Poisson statistics. In a time period $dt$, an expected number of comets $\lambda$ will generate an observed number of comets $k$ with probability
\begin{align}
P(k|\lambda)=\frac{\lambda^k e^{-\lambda}}{k!}.
\end{align}•
According to Bayes' theorem, the likelihood of a model $\lambda$ given the observed comet number $k$ will therefore be:
\begin{align}
P(\lambda|k)=\frac{P(k|\lambda)P(\lambda)}{P(k)}=\frac{\lambda^k e^{-\lambda}}{k!}\frac{P(\lambda)}{P(k)}
\end{align}•
where $P(\lambda)$ is the prior probability distributions for $\lambda$ and for $k$. For two different models $\lambda$ and $\lambda^\prime$ for this expected number, the likelihood ratio between these two models will therefore be given by
\begin{align}
\label{eq:att}
\mathcal{L}(\lambda,\lambda^\prime)=\left(\frac{\lambda}{\lambda^\prime}\right)^ke^{-(\lambda-\lambda^\prime)}\,\frac{P(\lambda)}{P(\lambda^\prime)}.
\end{align}•
In our case, we will be interested in the likelihood ratio between the dark matter explanation for the crater record vs. the constant rate explanation. The expected number of comet impacts in these two models will be
\begin{align}
\label{eq:model1}
\lambda&=dt\,r(t)\\
\label{eq:model2}
\lambda^\prime&=dt\, \frac{N}{T}.
\end{align}•
Here, Equation \ref{eq:model1} reflects our choice of the model where the comet rate is proportional to the instantaneous density $\rho(t)$, convolved with the Oort cloud response function $h(t)$:
\begin{align}
r(t)=\kappa [\,h\!*\!\rho\,](t).
\end{align}•
The constant $\kappa$ is fixed by demanding that the total integrated rate be equal to the total number of comets $N$. Equation \ref{eq:model2} simply gives the expected number of comets in time interval $dt$ for a constant rate giving total number of comets impacts $N$ in total time $T=250$ My. Likewise, the observed number of comets in time interval $dt$ is given by the crater record:
\begin{align}
k= dt\, C(t)
\end{align}•
where $C(t)$ is the crater rate history in events per My. It is a sum of Gaussian distributions for each crater with area 1, centered at its age and with width given by the uncertainty in its age.

Equation \ref{eq:att} defines the likelihood ratio at a given time $t$. To obtain the likelihood ratio over all times, we can assume the impacts probabilities at all times are independent and multiply the likelihood $L$ over all times. The log likelihood ratio over all times will therefore be given by the integral
\begin{align}
\log\mathcal{L}&=\sum_t k \log \left(\frac{\lambda}{\lambda^\prime}\right) + \lambda^\prime - \lambda +\log\left(\frac{P(\lambda)}{P(\lambda^\prime)}\right)\\
&=\int\! dt \, C(t) \log \left(\frac{r(t)}{N/T}\right) + \log\left(\frac{P(\Sigma_D)}{P(0)}\right).
\end{align}•
The $\lambda^\prime - \lambda$ term vanishes by the restriction that the total integrated rates are equal for the two models. $\Sigma_D$ is the surface mass density of the dark disk. $P(0)$ represents a model in which the dark disk surface density is zero. $P(\Sigma_D)/P(0)$ therefore represents the prior likelihood ratio between a model with and without a dark matter disk. We are thus comparing only two models: 1) a model where the comet rate is caused by solar oscillations with a dark matter disk of surface density $\Sigma_D$, and 2) a model in which $\Sigma_D=0$ and in which the comets arrive at a constant average rate $N/T$. The expression for the likelihood ratio is therefore
\begin{align}
\label{eq:likelihood}
\mathcal{L}=\exp\left\{\int\! dt \, C(t) \log \left(\frac{r(t)}{N/T}\right)\right\}\frac{P(\Sigma_D)}{P(0)}.
\end{align}•
For any given dark disk model $\Sigma_D$, we chose the thickness of the disk to be the minimum value permitted under the stability bound derived by \citet{shaviv16_2}. Under these assumptions, we find a posterior likelihood ratio of 10.

\subsection{Prior Probabilities}

\begin{figure}
\caption{The prior probabilities for a model with dark disk surface density $\Sigma_D$ and thickness given by the stability bound of \citet{shaviv16_2}.}
\plotone{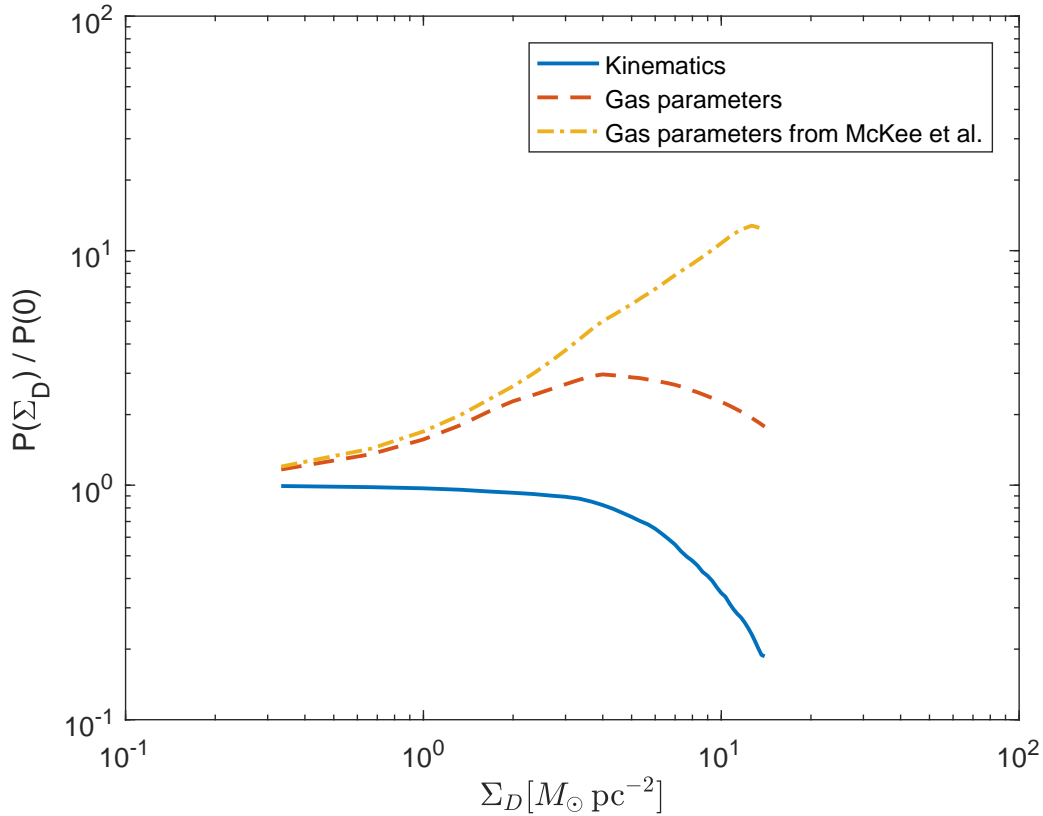}
\label{fig:priors}
\end{figure}

For the ratio of prior probabilities $P(\Sigma_D)/P(0)$ used in Equation \ref{eq:likelihood}, we rely on the results of \citet{paper1,paper2}. These contain a contribution from the stellar kinematics of A stars, taking into account the vertical epicyclic oscillations of these stars. The other contribution is from the gravithermal equilibirum of the Milky Way's interstellar gas. The values of $P(\Sigma_D)/P(0)$ are shown in Figure \ref{fig:priors} The priors inferred using the gas parameters of \citet{mckee} are also shown. As can be seen, the latter favor a higher value of $\Sigma_D$.

\subsection{Spiral Arm Crossing}
Another effect we account for here is the Sun's periodic crossing of the Milky Way's rotating spiral arm pattern. We assume spiral arms with a sinusoidal amplitude $A=0.25$  \citep{hessman}. This corresponds to an arm-interarm density ratio of 1.7 \citep[Ch.6]{bt}. As we are currently entering the Orion spur, we assume the current locally measured density to represent the azimuthally averaged density of the disk at the Solar radius. The Sun was assumed to move through the spiral arm pattern according the the ages in \citet{shaviv02}.

\subsection{Radial Oscillations}
For radial oscillations, epicycle theory \citep[Ch.3]{bt} gives an estimate of the mean radial position $R_g$ of the Sun:
\begin{equation}
\label{eq:epi}
R_\odot-R_g = \frac{V_\odot - V_c}{2B}
\end{equation}•
where $B=-12.37 \pm 0.64\, {\rm km s^{-1}\, kpc^{-1}}$ is one the Oort constants \citep{oortb}.
From \citet{schonrich}, we have
\begin{align}
\label{eq:vel}
(U_\odot,V_\odot-V_c,W_\odot)=(11.1 \pm 0.74, 12.24 \pm 0.47, 7.25 \pm0.37)\,{\rm km\,s^{-1}},
\end{align}•
This gives, according to Equation \ref{eq:epi}:
\begin{equation}
R_\odot-R_g = -0.49\pm0.03\, {\rm kpc}.
\end{equation}•
The radial oscillations will are assumed to vary with a frequency of $1.35\,\Omega_0$ \citep[Ch. 3]{bt}, where $\Omega_0=V_c/R_\odot$ is the angular velocity of the circular orbit at the position of the Sun. With the measured Oort constant $A-B=29.45 \pm 0.15\, {\rm km\, s^{-1} \,kpc^{-1}}$ \citep{Reid:2004rd} and Solar radius $R_\odot=8.33\pm 0.35\,{\rm kpc}$ \citep{gillesen}, we find the circular velocity at the Solar radius to be
\begin{align}
V_c= 245\pm10 {\rm km\, s^{-1}}.
\end{align}•
We therefore have
\begin{align}
1.35\,\Omega_0 = 1.35\,V_c/R_\odot = 39.7 \pm 2.3 \,{\rm km\,s^{-1}\,kpc^{-1} }
\end{align}•
Combining this  radial velocity $U_\odot$ of \ref{eq:vel}therefore gives a radial oscillation amplitude $0.57\pm0.03$ kpc and a present phase of $1.06\pm0.15$ rad. Assuming an exponential scale radius $R_{\rm MW}\simeq 3.5\pm0.5\, {\rm kpc}$ for the Galactic mass \citep{bt}, the radial oscillations will generate oscillations in density an amplitude of approximately $\simeq16.3\pm2.0\%$. The radial oscillations are therefore assumed to vary the density with an amplitude of 16.3\% and a period of 155 My.


\subsection{Solar Parameters}
Since there is a lot of uncertainty in the position of the Sun relative to the galactic plane, this will certainly affect whether the model can account for the dinosaur extinction. We therefore compute the likelihood for the value $Z_\odot = 26 \;{\rm pc}$ \citep{Z0cepheid} as well as for other values of $Z_\odot$. We also varied the vertical velocity of the Sun $W_0$ \citep{schonrich} and the dark halo density $\rho_{\rm halo}$ \citep{bovytr}.

\subsection{Stability Parameters}
The stability bound used here was that estimated by \citet{shaviv16_2}. For any dark disk surface density $\Sigma_D$, we estimated the disk's thickness $h_D(\Sigma_D)$ to be the minimum value consistent with the stability bound, i.e.
\begin{equation}
h_D = h_{D,{\rm min}}(\Sigma_D).
\end{equation}•
To estimate the effect of the uncertainty in this bound, we also used the values equal to half and double the value implied by the bound, i.e. $h_D=0.5$ or $1.5\, h_{D,{\rm min}}$.
%

\section{Results and Discussion}
\label{sec:results}

\begin{figure}
\caption{The solid line shows the likelihood ratio of oscillating Sun model relative to the constant
rate model. It shows a peak near $\Sigma_D=9\mppc{2}$. Dashed and notted lines show the posterior likelihood ratio after including the prior probabilities of Figure \ref{fig:priors}. The dot-dshed line shows the predicted comet rate at age
66 My as function of $\Sigma_D$.}
\plotone{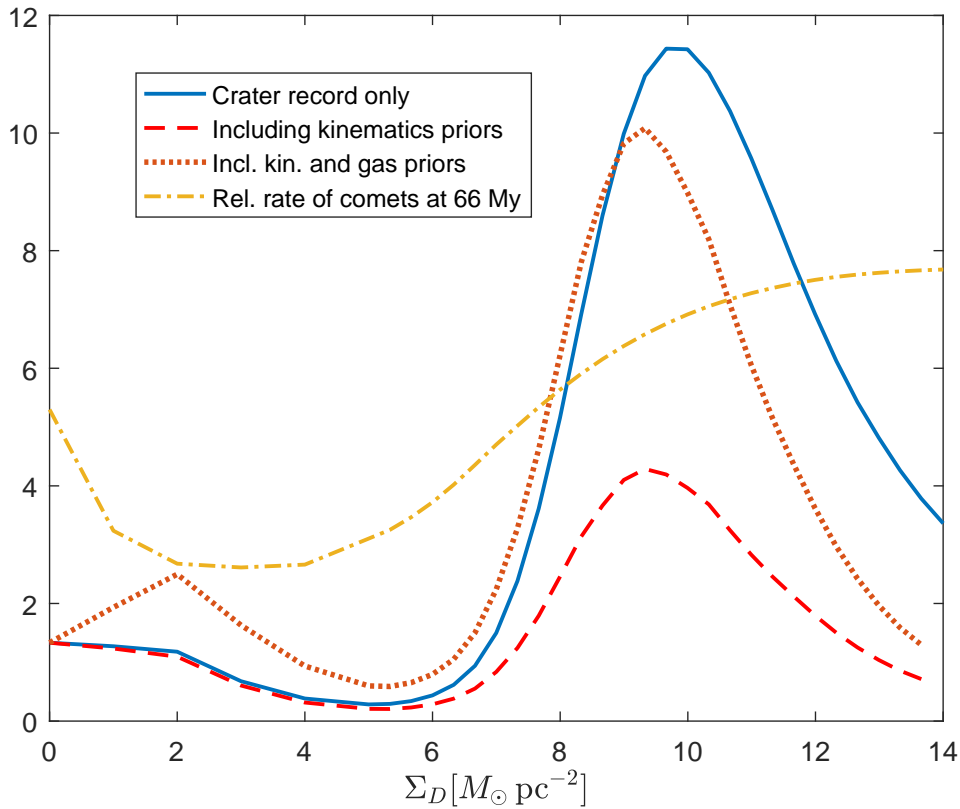}
\label{fig:likelihood}
\end{figure}

Figure \ref{fig:likelihood} shows the likelihood ratio for the two models as a function of dark disk surface density $\Sigma_D$, including the prior probabilities from Figure \ref{fig:priors}.The figure also shows the relative rate of comets at the date of 66 Mya as a function of $\Sigma_D$. For these calculations, the Galactic disk parameters for visible matter were taken from \citet{paper1}. The height of the Sun above the Galactic plane to be $Z_\odot=26\,{\rm pc}$, and a crater diameter cutoff of 20 km was imposed. We see that the likelihood ratio is as large as 10 in favor of an oscillatory model, at the best fit value $\Sigma_D\simeq 9\mppc{2}$. We also see that the relative rate of predicted comet impacts at age 66 My also is quite high near this value, indicating that the model correctly predicts the date of the 66 Mya Chicxulub crater.

\begin{figure}
\caption{The figure shows the history of craters larger than 20 km in diameter over the past 250 My as a probability density in age, as well as the predicted comet rate (in arbitrary units) assuming Solar oscillations with a best fit dark disk surface density of $\Sigma_D=9\mppc{2}$. Each blue spike represents a cratering shower. The recent craters of Popigai (36 Mya) and Chicxulub (66 Mya) are separated by only 29 My. The figure shows that the 50 My spent crossing the Carina / Sagittarius arm reduces the period for long enough to account for this smaller interval.}
\plotone{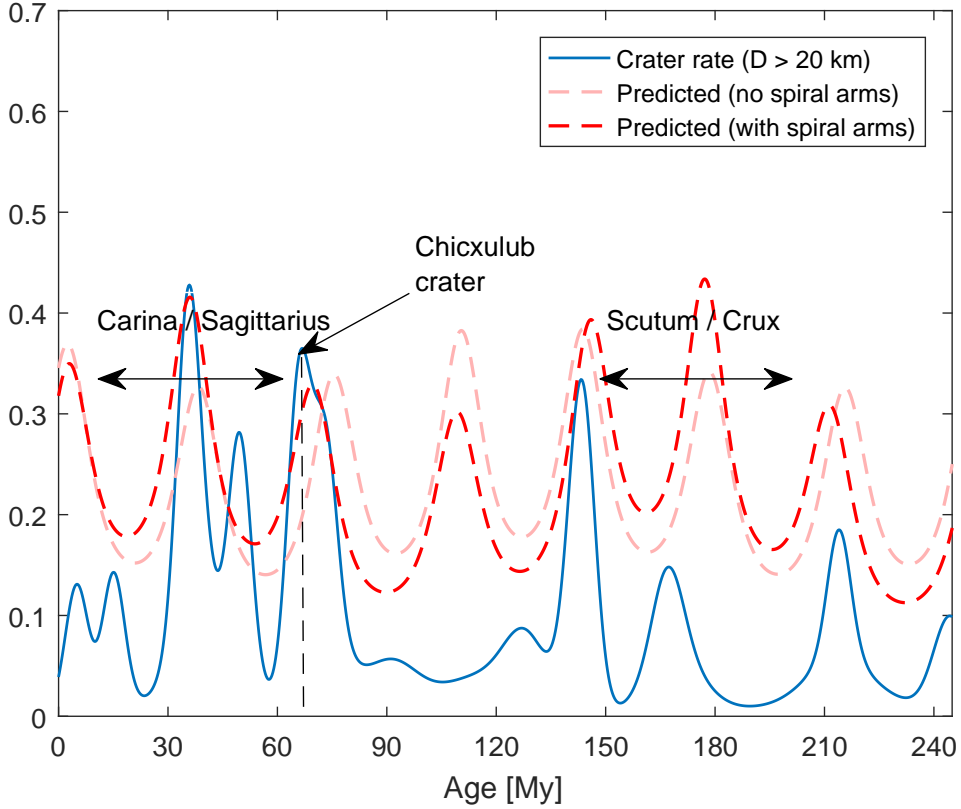}
\label{fig:cratersrate}
\end{figure}

Figure \ref{fig:cratersrate} shows the predicted comet rate in the oscillatory model superposed on the crater record. The time spent within spiral arms is shown by the double arrows. We can see that the the comet rate is consistent with the oscillations of the Sun when spiral arm crossing is taken into account. In particular, the Chicxulub crater matches the predicted rate.

\begin{figure}
\caption{Plots of likelihood ratio vs. $\Sigma_D$. The first plot on the left shows the effect of varying the parameters of the interstellar gas disk between those of \citet{paper1} and the right figure uses those of \citet{mckee}. The second plot shows the effect of varying the disk scale height relative to its minimum size $h_{\rm min}(\Sigma_D)$ under the stability bound of \citet{shaviv16_2}. The third plot and fourth plots show the effect of varying the present height of the Sun, $Z_\odot$ and its vertical velocity $W_\odot$.}
\plotone{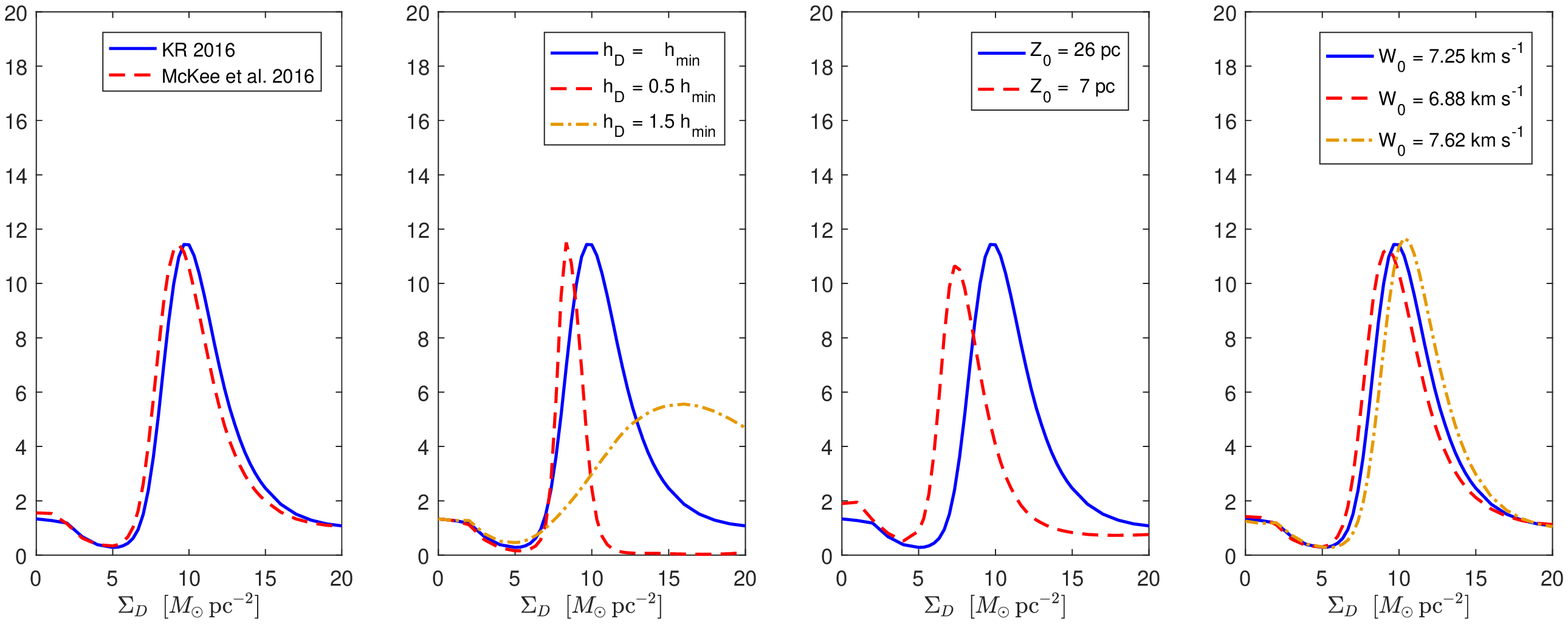}
\label{fig:plots}
\end{figure}

Figure \ref{fig:plots} shows the effect of varying the visible mass parameters, the dark disk scale height $h_D$, the present height of the Sun above the Galactic plane $Z_\odot$, the Sun's vertical velocity $W_0$, and the Sun's vertical velocity according to the ranges in Section \ref{sec:analysis}. We see that varying the visible mass parameters has a very minor effect on the results. Varying $h_D$ affects the best-fit value of $\Sigma_D$, with larger scale heights allowing more mass, as should be expected. lower values of $Z_\odot$ favor a lower value of $\Sigma_D$. Varying $W_\odot$ does not significantly affect the results.

\begin{figure}
\caption{We show here the inferred likelihood ratio as a function of dark disk surface density 1) including neither spiral arm crossing nor radial oscillations, 2) including radial oscillations but not spiral arms, and 3) including both. We can see that the radial oscillations decrease the likelihood of a dark disk model, while the spiral arms greatly enhance it.}
\plotone{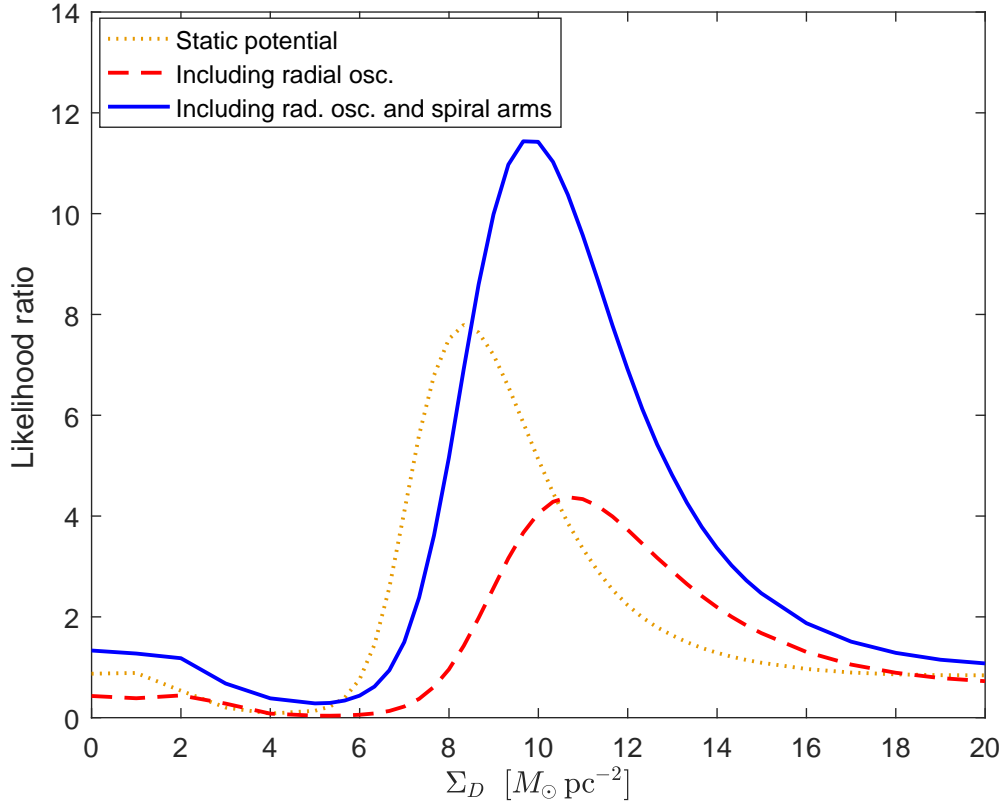}
\label{fig:norad}
\end{figure}

Figure \ref{fig:norad} shows the result of the computation1) including neither spiral arm crossing nor radial oscillations, 2) including radial oscillations but not spiral arms, and 3) including both. We can see that the radial oscillations decrease the likelihood of a dark disk model, while the spiral arms greatly enhance it. In both cases we see that although the data favors a higher dark disk surface density $\Sigma_D$.

\begin{figure}
\caption{Plots as in Figure \ref{fig:likelihood}. Here we vary the diameter cutoff between 20 km, 40 km, and 50 km respectively. Although the 20 km diameter cut may include objects that are not comets, a much higher significancance is obtained.}
\plotone{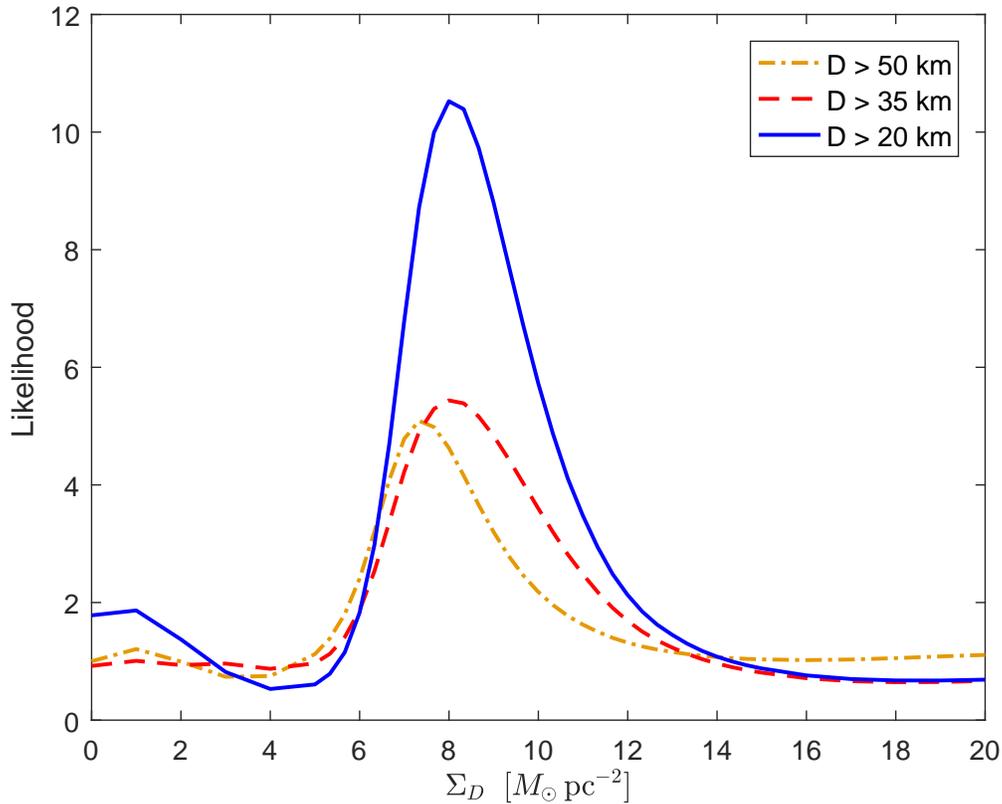}
\label{fig:diameters}
\end{figure}

Figure \ref{fig:diameters} shows the effect of varying the crater diameter cutoff. We find that the best fit values do not depend on the diameter cutoff, although a much higher significance is obtained for the 20 km cutoff, containing 26 craters.

\section{Conclusions}

We have shown that a dark disk consistent with stability bounds can indeed account for the observed comet rate on Earth. A dark disk model was found to be 10 times more likely than the constant rate model after prior probabilities were included. We have also shown that spiral arm crossings are crucial in accounting for variations in the comet rate. In particular, the cratering history on earth, including the date of the Chicxulub crater 66 Mya, is consistent with the predicted comet shower rate when the spiral arms are taken into account.

\acknowledgements

We would like to thank Lisa Randall and Matt Reece for helpful discussions. EDK was supported by NSF grants of Lisa Randall  PHY-0855591 and PHY-1216270. EDK and MR were supported by Harvard GSAS, Harvard Department of Physics, and Center for the Fundamental Laws of Nature. Calculations performed using MATLAB.

\end{document}